\documentclass[preprint,12pt]{elsarticle}

\usepackage{amssymb}
\usepackage{amsmath}

\begin{document}
\begin{frontmatter}

\title{Quantum no-key protocols for secret transmission of quantum and classical message}
\author[label1]{Li Yang}\ead{yang@is.ac.cn}
\author[label1]{Min Liang}
\author[label1]{Bao Li}
\author[label1]{Lei Hu}
\author[label2]{Ling-An Wu}
\address[label1]{State Key Laboratory of Information Security, Graduate University of Chinese Academy of Sciences,
Beijing 100049, China}
\address[label2]{Institute of Physics, Chinese Academy of Sciences, Beijing 100190, China}

\begin{abstract}
A theoretical framework of quantum no-key (QNK) protocol has been presented. As its applications, we develop three kinds of QNK protocols: the practical QNK protocols, the QNK protocol based on quantum perfect encryption, and the QNK protocols based on Boolean function computing. The security of these protocols is based on the laws of quantum mechanics, other than computational hypothesis.
\end{abstract}


\end{frontmatter}

\makeatletter
    \newcommand{\rmnum}[1]{\romannumeral #1}
    \newcommand{\Rmnum}[1]{\expandafter\@slowromancap\romannumeral #1@}
    \newcommand{\bm}[1]{\mbox{\boldmath{$#1$}}}
\makeatother

\section{Introduction}
The earliest group of quantum message oriented protocols is suggested in \cite{Boykin00, Ambainis00, Nayak07}, which can be regarded as a quantum version of one-time pad, the sender and the receiver must preshare secretly a classical key. Later, a public-key encryption scheme of quantum message is proposed \cite{Yang03}. Recently, this kind of public-key cryptosystems has been developed~\cite{Yang10}.

Here we consider another technique to securely transmit quantum message, so called quantum no-key (QNK) protocol. No-key protocol was first proposed by Shamir \cite{Menezes97}. It is a wonderful idea to transmit classical messages secretly in public channel, independent of the idea of public-key cryptosystem and that of secret-key cryptosystem. However, the protocol presented is computationally secure, cannot resists a man-in-the-middle(MIM) attack. \cite{Yangli02a,Yangli02} develop a quantum from of no-key protocol based on single-photon rotations, which can be used to transmit classical and quantum messages secretly. It can be seen that the security of the QNK protocol is based on the laws of quantum mechanics, so it is beyond computational hypothesis. \cite{Yangli03} proposed a protocol based on quantum computing of Boolean functions. This protocols is constructed with inherent identifications in order to prevent MIM attack. Similar to the idea of QNK protocol, Kanamori et al.\cite{Kanamori05} proposed a protocol for secure data communication, Kye et al.\cite{Kye05} proposed a quantum key distribution scheme, and Kak \cite{Subhash07} proposed a three-stage quantum cryptographic protocol for key agreement.\cite{Wu09} presents a practical QNK protocol, and studied a new kind of attack named unbalance-of-information-source (UIS) attack. This kind of attack may also be effective to quantum secure direct communication protocols, such as those in \cite{Beige01,Bostrom02,Deng2003,Deng2004}.

In this paper, we establish a theoretical framework of QNK protocol in Section \ref{sec:eqnkp}. Then we discuss some practical QNK protocols in Section \ref{sec:practical}. Based on quantum perfect encryption, we proposed a more general QNK protocol in Section \ref{sec:qnkp-qpe}. Finally, some protocols based on Boolean function computing are discussed in Section \ref{sec:qnkp-qc}.

\section{Essentials of quantum no-key protocol}\label{sec:eqnkp}
\subsection{Classical no-key protocol}
Shamir's no-key protocol \cite{Menezes97} is an encryption scheme to transmit messages without preshared keys. Assume encryption functions $E_A$ and $E_B$ are commutative, $E_B(E_A(*))=E_A(E_B(*))$. His idea is as follows:
\begin{enumerate}
\item{}
Alice encrypts the message $M$ with $k_A$ and sends Bob the message $C_1=E_A(M)$.

\item{}
Bob encrypts $C_1$ with $k_B$ and sends Alice the message $C_2=E_B(E_A(M))$.

\item{}
Alice decrypts $C_2$ through $D_A=(E_A)^{-1}$ and sends Bob $$C_3=D_A(E_B(E_A(M)))=D_A(E_A(E_B(M)))=E_B(M).$$

\item{}
Bob decrypts $C_3$ with $k_B$ to get $M$.
\end{enumerate}

The key point of this idea is that the two encryption functions $E_A$ and $E_B$ must be commutative,
\begin{equation}E_B(E_A(*))=E_A(E_B(*)).\end{equation}

\subsection{Some basic results relative to QNK protocol}
{\bf Lemma 1:} Operators $A$ and $B$ are unitary similar. If there exists unitary transformations $N$ and $M$ such that $NAM=B$, then $$\left[NP^{-1}\otimes(PM)^T-I\right]\overrightarrow{B}=0,P^{-1}BP=A;$$ or $$\left[N\otimes M^T-P^{-1}\otimes P^T\right]\overrightarrow{A}=0,P^{-1}AP=B;$$ where $P$ is unitary, $\overrightarrow{A}, \overrightarrow{B}$ are realignments of A, B, respectively.

{\bf Proof:} Operators $A$ and $B$ are unitary similar, so there exists unitary transformation $P$ satisfying $P^{-1}BP=A$. From $NAM=B$, it can be inferred that $NP^{-1}BPM=B$. Then we can conclude $[NP^{-1}\otimes(PM)^T]\overrightarrow{B}=\overrightarrow{B}$. That is $[NP^{-1}\otimes(PM)^T-I]\overrightarrow{B}=0$.

Operators $A$ and $B$ are unitary similar, so there exists unitary transformation $P'$ satisfying $P'^{-1}AP'=B$. From $NAM=B$, it can be inferred that $NAM=P'^{-1}AP'$. Then we can conclude $[N\otimes M^T]\overrightarrow{A}=[P'^{-1}\otimes P'^T]\overrightarrow{A}$. That is $[N\otimes M^T-P'^{-1}\otimes P'^T]\overrightarrow{A}=0$, where $P'^{-1}AP'=B$ and $P'$ is unitary.~$\Box$

{\bf Theorem 1:} Given four groups of operators $\{A_k,B_k,C_k,D_k|k=1,\cdots,d\}$, each group is a complete orthogonal basis of unitary operator space. If $D_lC_kB_lA_k=e^{i\varphi(k,l)}I,\forall k,l$, then
$$\left[e^{-i\varphi(k,l)}N\otimes M^T-A_k^{\dagger}\otimes A_k^T\right]\overrightarrow{B_l^\dagger}=0, \forall k,l,$$
where $M=C_kA_k$ is a unitary transformation only depending on $k$, and $N=D_lB_l$ is a unitary transformation only depending on $l$.

{\bf Proof:} Because $\{B_l|l=1,\cdots,d\}$ is a complete orthogonal basis of unitary operator space, there exists $\{\alpha_l\}$ satisfying $\sum_l\alpha_lB_l=I$.

Because $D_lC_kB_lA_k=e^{i\varphi(k,l)}I,\forall k,l$, and $D_l$ is unitary, it can be inferred that $C_kB_lA_k=e^{i\varphi(k,l)}D_l^\dagger, \forall k,l$. Then $$C_kA_k=C_k\left(\sum_l\alpha_lB_l\right)A_k=\sum_l\alpha_le^{i\varphi(k,l)}D_l^\dagger, \forall k.$$ Let $\sum_l\alpha_le^{i\varphi(k,l)}D_l^\dagger=M$, then $M=C_kA_k, \forall k$. Thus $M$ is a unitary transformation only depending on $k$.

In the same way, we can acquire $$D_lB_l=\sum_k\beta_ke^{i\varphi(k,l)}A_k^\dagger, \forall l.$$ Let $\sum_k\beta_ke^{i\varphi(k,l)}A_k^\dagger=N$, then $N=D_lB_l, \forall l$. Thus $N$ is a unitary transformation only depending on $l$.

From $M=C_kA_k$ and $N=D_lB_l$, it can be concluded that $C_k=MA_k^\dagger$,$D_l=NB_l^\dagger$. Because $D_lC_kB_lA_k=e^{i\varphi(k,l)}I$, one can obtain $NB_l^\dagger MA_k^\dagger B_lA_k=e^{i\varphi(k,l)}I$, so $e^{-i\varphi(k,l)}NB_l^\dagger M=A_k^\dagger B_l^\dagger A_k$. Because $B_l^\dagger$ and $A_k^\dagger B_l^\dagger A_k$ are unitary similar, one can conclude from the Lemma 1 that $$\left[e^{-i\varphi(k,l)}N\otimes M^T-A_k^\dagger\otimes A_k^T\right]\overrightarrow{B_l^\dagger}=0, \forall k,l.$$~$\Box$

{\bf Theorem 2:} Suppose $\{A_k,B_k,C_k,D_k|k=1,\cdots,d\}$ satisfy the conditions in Theorem 1, and $C_k=A_k^\dagger$, $D_k=B_k^\dagger$, $\forall k$. Then $C_kB_l=e^{i\varphi(k,l)}B_lC_k$ is sufficient and necessary for $D_lC_kB_lA_k=e^{i\varphi(k,l)}I,\forall k,l$.

{\bf Proof:}
(sufficient) From $C_kB_l=e^{i\varphi(k,l)}B_lC_k$, we can know $D_lC_kB_lA_k=e^{i\varphi(k,l)}D_lB_lC_kA_k$. Because $C_k=A_k^\dagger$, $D_l=B_l^\dagger$, then $D_lC_kB_lA_k=e^{i\varphi(k,l)}I$.

(necessary) From $C_k=A_k^\dagger$ and $D_k=B_k^\dagger$, $\forall k$, we know that $M=N=I$ and $NM=D_lB_lC_kA_k=I$. Because $D_lC_kB_lA_k=e^{i\varphi(k,l)}I$, $D_lC_kB_lA_k=e^{i\varphi(k,l)}D_lB_lC_kA_k$. Then $C_kB_l=e^{i\varphi(k,l)}B_lC_k$.~$\Box$

\subsection{Quantum commutative transformation and QNK protocol}
Usually, we call two quantum transformation $U_A$ and $U_B$ are commutative if $U_AU_B=U_BU_A$. Sometimes in this paper we prefer an entended definition: $U_AU_B=e^{i\varphi}U_BU_A$. Similar to commutative algorithm in Shamir's classical no-key protocol, quantum commutative transformations are used to construct QNK protocol.

Let $\{U_{A_i}\}$ and $\{U_{B_i}\}$ are two sets of unitary operations, we suppose each pair of $U_{A_i}$ and $U_{B_j}$ are commutative. The QNK protocol is as follows:
\begin{enumerate}
\item{}
Alice randomly selects a number $i$, and encrypts quantum state $\rho$ with $U_{A_i}$, and sends Bob $\rho_1=U_{A_i}\rho U_{A_i}^{\dagger}$.
\item{}
Bob randomly selects a number $j$, and encrypts $\rho_1$ with $U_{B_j}$ and sends Alice $\rho_2=U_{B_j}\rho_1 U_{B_j}^{\dagger}=U_{B_j}U_{A_i}\rho U_{A_i}^{\dagger}U_{B_j}^{\dagger}$.
\item{}
Alice decrypts $\rho_2$ with $U_{A_i}^{\dagger}$ and sends Bob $\rho_3=U_{A_i}^{\dagger}\rho_2U_{A_i}= \\ U_{A_i}^{\dagger}U_{B_j}U_{A_i}\rho U_{A_i}^{\dagger}U_{B_j}^{\dagger}U_{A_i}=U_{A_i}^{\dagger}U_{A_i}U_{B_j}\rho U_{B_j}^{\dagger}U_{A_i}^{\dagger}U_{A_i}=U_{B_j}\rho U_{B_j}^{\dagger}$.
\item{}
Bob decrypts $\rho_3$ with $U_{B_j}^{\dagger}$, and gets $U_{B_j}^{\dagger}\rho_3U_{B_j}=\rho$.
\end{enumerate}

{\bf Proposition 1:} Suppose both $U_A$ and $U_B$ are unitary transformations. Then the three conditions $U_AU_B=U_BU_A$, $U_B^\dagger U_A^\dagger=U_A^\dagger U_B^\dagger$ and $U_BU_A^\dagger=U_A^\dagger U_B$ are equivalent.

{\bf Proof:} It can be seen that, if $U_A$ and $U_B$ satisfies any one of the following conditions:
\begin{align}
& U_AU_B=U_BU_A,\label{eq1} \\
& U_B^\dagger U_A^\dagger=U_A^\dagger U_B^\dagger,\label{eq2} \\
& U_BU_A^\dagger=U_A^\dagger U_B,\label{eq3}
\end{align}
then $U_B^\dagger U_A^\dagger U_BU_A=I$ holds. Because $U_A$ and $U_B$ are unitary transformations, $U_A^\dagger U_A=I$ and $U_B^\dagger U_B=I$. From the identity $U_B^\dagger U_A^\dagger U_BU_A=I$, we can deduce all of the above three identities (\ref{eq1}),(\ref{eq2}),(\ref{eq3}). Thus $U_B^\dagger U_A^\dagger U_BU_A=I$ is equivalent with any one of the three identities. This means the three conditions are equivalent.~$\Box$

{\bf Remark 1:} Three instances of quantum commutative transformation are as follows:
\begin{enumerate}
\item{}
Making a transformation directly on the bases:
\begin{equation*}U_A(\sum_m\alpha_m|m\rangle)=\sum_m\alpha_m|m\oplus s_A\rangle,\end{equation*}
\begin{equation*}U_B(\sum_m\alpha_m|m\rangle)=\sum_m\alpha_m|m\oplus s_B\rangle,\end{equation*}

\item{}
Making use of an auxiliary register:
\begin{equation*}U_A(\sum_m\alpha_m|m\rangle|s\rangle)=\sum_m\alpha_m|m\rangle|s\oplus F_A(m)\rangle,\end{equation*}
\begin{equation*}U_B(\sum_m\alpha_m|m\rangle|s\rangle)=\sum_m\alpha_m|m\rangle|s\oplus F_B(m)\rangle,\end{equation*}

\item{}
Making use of two auxiliary registers:
\begin{equation*}U_A(\sum_m\alpha_m|m\rangle|0\rangle|0\rangle)=\sum_m\alpha_m|m\rangle|F_A(m)\rangle|0\rangle,\end{equation*}
\begin{equation*}U_B(\sum_m\alpha_m|m\rangle|0\rangle|0\rangle)=\sum_m\alpha_m|m\rangle|0\rangle|F_B(m)\rangle,\end{equation*}
\end{enumerate}

{\bf Remark 2:} The protocol in this section does not have inherent identification and cannot resistant man-in-the-middle attack. For example, if Eve intercepts $\rho_1$, she does nothing before sends it back to Alice, Alice decrypts $\rho_1$ with $U_{A_i}^{\dagger}$ and sends $\rho$, thus Eve can obtain the message $\rho$. Therefore, we have to construct QNK protocol with personal identification.

\subsection{Theoretical framework of quantum no-key protocol}
Quantum message space is denoted as $H_M$. Two sets of pair operators $\{U_k,U_k'\}$ and $\{V_l,V_l'\}$ are two public sets of unitary operators which performs on $H_M$, where $k,l\in\{0,1,\cdots, d\}$. Alice uses the set $\{U_k,U_k'|k\in\{0,1,\cdots, d\}\}$, while Bob uses the set $\{V_l,V_l'|l\in\{0,1,\cdots, d\}\}$. Suppose Alice wants to send quantum message $\rho\in H_M$. The framework of quantum no-key protocol is as follows:
\begin{enumerate}
\item{}
Alice randomly selects a number $k\in\{0,1,\cdots, d\}$, then performs $U_k$ on the quantum message $\rho$, and gets $\rho_1=U_k\rho U_k^\dagger$. Then she sends $\rho_1$ to Bob.
\item{}
Bob receives the message $\rho_1$, then randomly selects $l\in\{0,1,\cdots, d\}$. He performs $V_l$ on $\rho_1$, and gets $\rho_2=V_l\rho_1 V_l^\dagger$. Then he sends $\rho_2$ to Alice.
\item{}
Alice receives $\rho_2$, then performs $U_k'$ on $\rho_2$ and gets $\rho_3=U_k'\rho_2 U_k'^\dagger$. Then she sends $\rho_3$ to Bob.
\item{}
Bob receives $\rho_3$, then performs $V_l'$ on $\rho_3$ and gets $\rho=V_l'\rho_3V_l'^\dagger$.
\end{enumerate}

Note that the number $k$ and $l$ are selected from two independent uniform distributions.

{\bf Proposition 2:} The protocol holds if and only if $V_l'U_k'V_lU_k=e^{i\varphi(k,l)}I$, $\forall k,l\in\{0,1,\cdots, d\}$.

{\bf Proof:} It is obvious that the protocol holds if and only if $$\rho=V_l'U_k'V_lU_k\rho U_k^\dagger V_l^\dagger U_k'^\dagger V_l'^\dagger, \forall \rho\in H_M, \forall k,l\in\{0,1,\cdots, d\}$$. That means, the protocol holds if and only if $V_l'U_k'V_lU_k=e^{i\varphi(k,l)}I$, $\forall k,l\in\{0,1,\cdots, d\}$.~$\Box$

According to Theorem 1, we conclude from $V_l'U_k'V_lU_k=e^{i\varphi(k,l)}I$ that
\begin{equation}\label{eq4}
\left[e^{-i\varphi(k,l)}N\otimes M^T-U_k^\dagger\otimes U_k^T\right]\overrightarrow{V_l^\dagger}=0, \forall k,l\in\{0,1,\cdots, d\},
\end{equation}
where $M=U_k'U_k$ is a unitary transformation only depending on $k$, and $N=V_l'V_l$ is a unitary transformation only depending on $l$. Thus, the following proposition holds.

{\bf Proposition 3:} Eq.(\ref{eq4}) is a necessary condition for the protocol holding.~$\Box$

Let us consider a special case of $U_k'=U_k^\dagger$, $V_l'=V_l^\dagger$. According to Theorem 2 and Proposition 2, we can infer that

{\bf Proposition 4:} Suppose the conditions $U_k'=U_k^\dagger$, $V_l'=V_l^\dagger$ are satisfied, then the protocol holds if and only if $U_k^\dagger V_l=e^{i\varphi(k,l)}V_lU_k^\dagger$, $\forall k,l\in\{0,1,\cdots, d\}$.~$\Box$

Let us consider a more general framework of quantum no-key protocol, in which two ancillary states are used. Suppose Alice will send quantum message $\rho\in H_M$. The ancillary states used by Alice and Bob are $\rho_A$ and $\rho_B$, respectively. The framework of QNK protocol is described as (see Figure~\ref{fig1}):

\begin{figure}[htp!]
\begin{center}
\includegraphics[width=12cm]{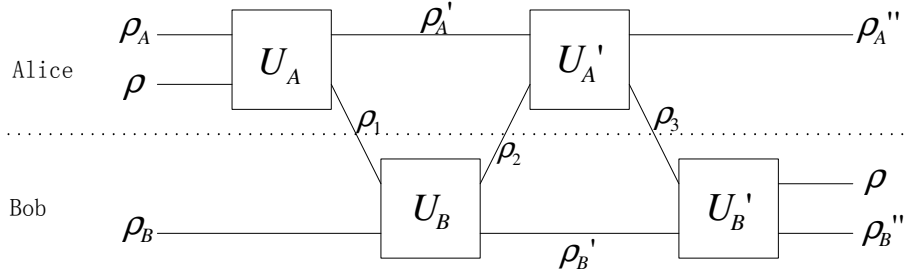}
\end{center}
\caption{\label{fig1} A general framework of quantum no key protocol. This figure is divided into two part by a dashed line. The part above the dashed line describes Alice's operations, and the other part describes Bob's operations. The quantum state $\rho$ is the plain state, and $\rho_1,\rho_2,\rho_3$ represents the three cipher states transmitted between Alice and Bob. $\rho_A,\rho_B$ are two ancillary states generated randomly by Alice and Bob, respectively.}
\end{figure}

\begin{enumerate}
\item{}
Alice randomly prepare a quantum state $\rho_A$, then performs $U_A$ on the quantum states $\rho_A\otimes\rho$ and gets $U_A(\rho_A\otimes\rho)U_A^\dagger$. Then she sends to Bob the first cipher state $\rho_1$, $$\rho_1=tr_A(U_A(\rho_A\otimes\rho)U_A^\dagger)\triangleq \mathcal{E}_A(\rho).$$
She retains the state $\rho_A'=tr_M(U_A(\rho_A\otimes\rho)U_A^\dagger)$.
\item{}
Bob randomly prepares a quantum state $\rho_B$, then performs $U_B$ on the quantum states $\rho_1\otimes\rho_B$  and gets $U_B(\rho_1\otimes\rho_B)U_B^\dagger$. Then he sends to Alice the second cipher state $\rho_2$, $$\rho_2=tr_B(U_B(\rho_1\otimes\rho_B)U_B^\dagger)\triangleq \mathcal{E}_B(\rho_1).$$
He retains the state $\rho_B'=tr_M(U_B(\rho_1\otimes\rho_B)U_B^\dagger)$.
\item{}
Alice performs $U_A'$ on $\rho_A'\otimes\rho_2$, and sends to Bob the third cipher state $\rho_3$, $$\rho_3=tr_A(U_A'(\rho_A'\otimes\rho_2)U_A'^\dagger)\triangleq\mathcal{E}_A'(\rho_2).$$
\item{}
Bob performs $U_B'$ on $\rho_3\otimes\rho_B'$, and gets the message $\rho\prime$, $$\rho'=e^{i\phi}\rho=tr_B(U_B'(\rho_3\otimes\rho_B')U_B'^\dagger)\triangleq\mathcal{E}_B'(\rho_3).$$
\end{enumerate}

This protocol holds if and only if the four quantum operations satisfy the condition
\begin{equation}
\mathcal{E}_B'\circ\mathcal{E}_A'\circ\mathcal{E}_B\circ\mathcal{E}_A=e^{i\phi}\mathcal{I}.
\end{equation}

As a special case, the unitary transformations $U_A$,$U_B$ can be chosen as bitwise controlled-unitary transformations where the message qubits act as control qubits, and $U_A'=U_A^\dagger$,$U_B'=U_B^\dagger$. In this case, $(I\otimes U_B)(U_A \otimes I)=(U_A \otimes I)(I\otimes U_B)$, and $(I\otimes U_B')(U_A' \otimes I)(I\otimes U_B)(U_A \otimes I)=I$.

\subsection{Quantum no-key protocol with personal identification}\label{sec:qnkp-pi}
Denote quantum message space as $H_M$, identification space as $H_A$. Alice and Bob preshare an identification key $(s_A,s_B)$. The protocol is as follows:
\begin{enumerate}
\item{}
Alice randomly selects a number $k\in\{0,1,\cdots, d\}$, then performs $U_k(s_A)$ on the quantum message $\rho\in H_M$ associated with ancillary qubits $|0\rangle\langle0|\in H_A$, and gets $\rho_1=U_k(s_A)\left(\rho\otimes|0\rangle\langle0|\right)U_k(s_A)^\dagger$ $\in H_M\otimes H_A$. Then she sends $\rho_1$ to Bob.
\item{}
Bob receives the message $\rho_1$, then randomly selects $l\in\{0,1,\cdots, d\}$, performs $V_l'(s_A)$ on $\rho_1$ and measures the ancillary qubits (Here it is required that $V_l'(s_A)$ satisfies $V_l'(s_A)\rho_1V_l'(s_A)^\dagger=\rho_1'\otimes|0\rangle\langle0|$). After measurement, the message collapses to $\rho_1'\in H_M$. He admits $\rho_1$ comes from Alice if the result of measurement is $0$. While passing the identification, he uses $V_l''(s_B)$ to compute $\rho_2=V_l''(s_B)\left(\rho_1'\otimes|0\rangle\langle0|\right)V_l''(s_B)^\dagger$ $\in H_M\otimes H_A$, and sends $\rho_2$ to Alice.
\item{}
Alice receives $\rho_2$, then performs $U_k'(s_B)$ on $\rho_2$ and measures the ancillary qubits. She admits $\rho_2$ comes from Bob if the result of measurement is $0$. After that, she uses $U_k''(s_A)$ to compute $\rho_3=U_k''(s_A)\left(\rho_2'\otimes|0\rangle\langle0|\right)U_k''(s_A)^\dagger$ $\in H_M\otimes H_A$, and sends $\rho_3$ to Bob.
\item{}
Bob receives $\rho_3$, then performs $V_l(s_A)$ on $\rho_3$ and measures the ancillary qubits. He admits $\rho_3$ comes from Alice if the result of measurement is $0$. After measurement, the message collapses to quantum message $\rho\in H_M$.
\end{enumerate}

In this protocol, operators $U_k(s),U_k'(s),U_k''(s),V_l(s),V_l'(s),V_l''(s)$ are unitary transformations performing on the whole space $H_M\otimes H_A$. The protocol is correct if and only if the following conditions hold: $\forall s_A,s_B,k,l$,
\begin{align}
& V_l'(s_A)U_k(s_A)\left(\rho\otimes|0\rangle\langle 0|\right)U_k(s_A)^\dagger V_l'(s_A)^\dagger=\rho_1'\otimes |0\rangle\langle 0|,\\
& U_k'(s_B)V_l''(s_B)\left(\rho_1'\otimes|0\rangle\langle 0|\right)V_l''(s_B)^\dagger U_k'(s_B)^\dagger=\rho_2'\otimes |0\rangle\langle 0|,\\
& V_l(s_A)U_k''(s_A)\left(\rho_2'\otimes|0\rangle\langle 0|\right)U_k''(s_A)^\dagger V_l(s_A)^\dagger=\rho\otimes |0\rangle\langle 0|.
\end{align}

Furthermore, these three equations are equivalent to the following conditions:
\begin{align}
& V_l'(s_A)U_k(s_A)=U_M(k,l,s_A)\otimes I_A, \forall k,l,s_A,\label{eq5}\\
& U_k'(s_B)V_l''(s_B)=U_M'(k,l,s_B)\otimes I_A, \forall k,l,s_B,\label{eq6}\\
& V_l(s_A)U_k''(s_A)=U_M''(k,l,s_A)\otimes I_A, \forall k,l,s_A,\label{eq7}
\end{align}
where $U_M(k,l,s_A)$, $U_M'(k,l,s_B)$, $U_M''(k,l,s_A)$ are unitary operators performing on $H_M$ and satisfy the relation $U_M''(k,l,s_A)U_M'(k,l,s_B)U_M(k,l,s_A)=I_M$, $\forall s_A,s_B,k,l$.

The preshared key $(s_A,s_B)$ are used for $3$ times to identify each other. If we require the quantum state obtained after each measurement be independent with the identification key $(s_A,s_B)$, that means $\rho_1'$, $\rho_2'$ and $\rho$ are independent with $s_A$ and $s_B$, thus $U_M(k,l,s_A)$, $U_M'(k,l,s_B)$, $U_M''(k,l,s_A)$ are also independent with $s_A$ and $s_B$, the Eq.(\ref{eq5})(\ref{eq6})(\ref{eq7}) can be written as follows:
\begin{align}
& V_l'(s_A)U_k(s_A)=U_M(k,l)\otimes I_A, \forall k,l,s_A,\\
& U_k'(s_B)V_l''(s_B)=U_M'(k,l)\otimes I_A, \forall k,l,s_B,\\
& V_l(s_A)U_k''(s_A)=U_M''(k,l)\otimes I_A, \forall k,l,s_A,
\end{align}
where $U_M''(k,l)U_M'(k,l)U_M(k,l)=I_M$, $\forall k,l$.

\section{Practical quantum no-key protocol}\label{sec:practical}

The protocols in this section are based on rotation of single photon, and may be implemented with current technology.

Generally speaking, two rotation transformations on the Bloch sphere are not commutative, unless the axes are parallel. Thus the key technique of this protocol is that Bob's encryption rotation and Alice's decryption rotation must be commutative. It can be proven that in this case the two axes of rotations must be parallel.

{\bf Proposition 5:} The rotation transformations on the sphere are commutative if and only if the axes are parallel.

{\bf Proof:} Denote two axes are ${\bf{n}}_1$ and ${\bf{n}}_2$, and the rotation transformations $U_{{\bf{n}_1}}$$(\varphi_1)$ and $U_{{\bf{n}}_2}(\varphi_2)$ represents the rotation around the axes ${\bf{n}}_1$ and ${\bf{n}}_2$ by an angle $\phi_1$ and an angle $\phi_2$, respectively. Because
\begin{equation}
({\bf{n}}_1\cdot{\bm \sigma})({\bf{n}}_2\cdot{\bm \sigma})={\bf{n}}_1\cdot{\bf{n}}_2+i{\bm \sigma}\cdot({\bf{n}}_1\times{\bf{n}}_2),
\end{equation}
therefore
\begin{equation}
\left[{\bf{n}}_1\cdot{\bm \sigma},{\bf{n}}_2\cdot{\bm \sigma}\right]=2i({\bf{n}}_1\times{\bf{n}}_2)\cdot{\bm \sigma}.
\end{equation}
For the rotation operator
\begin{equation}
U_{\bf{n}}(\varphi)=\textrm{exp}\left(\frac{1}{2}i\varphi\bf{n}\cdot{\bm \sigma}\right)=\textrm{cos}\frac{1}{2}\varphi+i{\bf{n}}\cdot{\bm \sigma}\textrm{sin}\frac{1}{2}\varphi,
\end{equation}
we have
\begin{equation}
\left[U_{{\bf{n}}_1}(\varphi_1),U_{{\bf{n}}_2}(\varphi_2)\right]=-2i\textrm{sin}\frac{1}{2}\varphi_1\textrm{sin}\frac{1}{2}\varphi_2({\bf{n}}_1\times{\bf{n}}_2)\cdot{\bm \sigma}.
\end{equation}

Suppose that both rotations are non-zero, then the two rotations are commutative if and only if the two axes are parallel.~$\Box$

\subsection{Protocol for quantum message transmission\cite{Yangli02}}\label{sec:protocol-qmt}
Let us consider the secret transmission of a quantum message in product state. Denote $U_{\bf{n}}(\varphi)$ as a rotation around axis $\bf{n}$ by an angle $\varphi$.
In Bloch sphere representation, the state of a qubit can be denoted as $|\bf{n},\varphi\rangle$, which can be prepared using a rotation operator $U_{\bf{n}}(\varphi)$, $|\bf{n},\varphi\rangle=$ ${U}_{\bf{n}}(\varphi)$ $|0\rangle$.
The protocol is as follows:
\begin{enumerate}
\item{}
Alice chooses $m$ qubits for transformation:
\begin{equation}\label{quantum-messgae}|{\bf{n}}_{10},\varphi_1\rangle,\cdots,|{\bf{n}}_{m0},\varphi_m\rangle.\end{equation}
\item{}
Alice chooses $\varphi_{A_i}(i=1,2,\cdots,m)$ randomly from a $K$-element set
\begin{equation}\label{K-element}\{\alpha_k=\frac{k\pi}{K}|k=0,1,\cdots,2K-1\}.\end{equation}
\item{}
Alice chooses randomly ${\bf{n}}_i$(i=1,2,$\cdots,$m), and opens them.
\item{}
Alice prepares $m$ single-photons, with the $i$-th photon in the state
\begin{equation}|\Psi_i\rangle_{A_1}=U_{{\bf{n}}_i}(\varphi_{A_i})|{\bf{n}}_{i0},\varphi_i\rangle,\end{equation}
then sends these photons to Bob one by one.

\item{}
Bob chooses $\varphi_{B_i}(i=1,2,\cdots,m)$ randomly from the $K$-element set (\ref{K-element}) by means of local random number source, and changes the polarization directions of photons separately as below:
\begin{equation}|\Psi_i\rangle_{B_1}=U_{{\bf{n}}_i}(\varphi_{B_i})U_{{\bf{n}}_i}(\varphi_{A_i})|{\bf{n}}_{i0},\varphi_i\rangle,\end{equation}
then sends back these photons to Alice.

\item{}
Alice removes her encryption transformation of the photons and gets
\begin{equation}|\Psi_i\rangle_{A_2}=U_{{\bf{n}}_i}(\varphi_{B_i})|{\bf{n}}_{i0},\varphi_i\rangle,\end{equation}
then sends them to Bob again.

\item{}
Bob removes his encryption transformation of the photons and gets
\begin{equation}|\Psi_i\rangle_{B_2}=|{\bf{n}}_{i0},\varphi_i\rangle,\end{equation}
then he gets the message (\ref{quantum-messgae}).
\end{enumerate}

Because $\varphi_{A_i}$,$\varphi_{B_i}$ are chosen from set (\ref{K-element}) randomly and independently, Eve cannot get any information from simple intercept/resend attack. Unfortunately, These two protocols cannot defend MIM (of quantum channel only) attack, even through there is an authenticated classical channel.

{\bf Remark 3:} The quantum state in the protocol should be written in the form of density matrix.
However, for understanding easily, the quantum states are written in the form of ware function instead of density matrix, whenever making no confusion. We can rewrite the above protocol in the following form:
\begin{enumerate}
\item{}
Alice chooses $m$ photons in this quantum state
$$
\rho=|{\bf{n}}_1,\varphi_1\rangle\langle{\bf{n}}_1,\varphi_1|\otimes\cdots\otimes|{\bf{n}}_m,\varphi_m\rangle\langle{\bf{n}}_m,\varphi_m|.
$$

\item{}
Alice performs $m$-qubit rotation $$U_{A}=U_{{\bf{n}}_1}(\varphi_{A_1})\otimes\cdots\otimes U_{{\bf{n}}_m}(\varphi_{A_m})$$ on the $m$ qubits, and get the state $$\rho_1=\rho_{A_1}\otimes\cdots\otimes\rho_{A_m}$$
where $\rho_{A_i}=U_{{\bf{n}}_i}(\varphi_{A_i})|{\bf{n}}_{i0},\varphi_i\rangle\langle{\bf{n}}_{i0},\varphi_i|U_{{\bf{n}}_i}^\dagger(\varphi_{A_i})$,
then sends these photons to Bob one by one.

\item{}
Bob performs $m$-qubit rotation $$U_{B}=U_{{\bf{n}}_1}(\varphi_{B_1})\otimes\cdots\otimes U_{{\bf{n}}_m}(\varphi_{B_m})$$ on the state $\rho_1$ and get the state $$\rho_2=\rho_{B_1}\otimes\cdots\otimes\rho_{B_m}$$
where $\rho_{B_i}=U_{{\bf{n}}_i}(\varphi_{B_i})U_{{\bf{n}}_i}(\varphi_{A_i})|{\bf{n}}_{i0},\varphi_i\rangle\langle{\bf{n}}_{i0},\varphi_i|U_{{\bf{n}}_i}^\dagger(\varphi_{A_i})U_{{\bf{n}}_i}^\dagger(\varphi_{B_i})$,
then sends back these photons to Alice.

\item{}
Alice receives these qubits and removes her rotations on the qubits by performing rotation $$U_{A}^\dagger=U_{{\bf{n}}_1}(-\varphi_{A_1})\otimes\cdots\otimes U_{{\bf{n}}_m}(-\varphi_{A_m})$$ on the $m$ qubits, and then gets the state $$\rho_3=\rho_{A_1}'\otimes\cdots\otimes\rho_{A_m}'$$
where $\rho_{A_i}'=U_{{\bf{n}}_i}(\varphi_{B_i})|{\bf{n}}_{i0},\varphi_i\rangle\langle{\bf{n}}_{i0},\varphi_i|U^\dagger_{{\bf{n}}_i}(\varphi_{B_i})$,
then sends them to Bob again.

\item{}
Bob receives these qubits and removes her rotations on the qubits by performing rotation $$U_{B}^\dagger=U_{{\bf{n}}_1}(-\varphi_{B_1})\otimes\cdots\otimes U_{{\bf{n}}_m}(-\varphi_{B_m})$$ on the $m$ qubits. Since $U_A$ and $U_B$ are commutative, Bob can get the message $\rho$
\end{enumerate}

It can be seen that, according to Proposition 5, $U_A$ and $U_B$ are commutative if and only if the axes of rotations $U_{A_i}(\varphi_{A_i})$ and $U_{B_i}(\varphi_{B_i})$ are parallel for every $i$.

\subsection{Protocol with personal identification\cite{Yangli02}}\label{sec:protocol}
Personal identification is necessary to defend MIM attack. We modified the protocol in Section \ref{sec:protocol-qmt} as following:

Alice and Bob share $\{\varphi_{C_i}|i=1,\cdots,n\}$ secretly before communication. In the second step, Alice rotates each photon by an angle $\varphi_{C_i}(i=1,\cdots,n)$, then sends $|\varphi_i+\varphi_{A_i}+\varphi_{C_i}\rangle(i=1,\cdots,n)$ to Bob. It continues according to the original protocol and Bob will get the states $|\varphi_i+\varphi_{B_i}+\varphi_{C_i}\rangle(i=1,\cdots,n)$ in the fourth step. Because Bob knows the value of $\varphi_{C_i}$ and $\varphi_{B_i}$, he can remove them and get the quantum message (\ref{quantum-messgae}).

The authentication information $\{\varphi_{C_i}|i=1,\cdots,n\}$ can be used repeatedly under the protection of continuously changed local random numbers $\{\varphi_{A_i},\varphi_{B_i}|i=1,\cdots,n\}$.

\subsection{Practical scheme with mutual identification\cite{Wu09}}\label{ps-mi}
In \cite{Wu09}, we proposed a quantum no-key protocol with mutual identification. In this protocol, the photons are transmitted group by group. In each group, there are $n+m$ photons. $n$ photons are used to transmit information, called IF-photons, $m$ photons are used for identification, called ID-photons. The protocol is as follows:
\begin{enumerate}
\item{}
Alice operates IF-photons and ID-photons differently. For the $j$-th ID-photons: Alice prepares $|\psi_j\rangle_{P_j}$, where $P_j$ is the position of the $j$-th ID-photon. For the $i$-th IF-photon: Alice prepares $|\varphi_{S_i}+\varphi_{A_i}+\varphi_{C_i}\rangle_{Q_i}$, where $\varphi_{A_i}$ is a random angle and $Q_i$ is the position of the $i$-th IF-photon. At last Alice sends the first message to Bob.

\item{}
For $i$-th IF-photon: Bob uses $R(\varphi_{B_i})$ to get state $|\varphi_{S_i}+\varphi_{A_i}+\varphi_{C_i}+\varphi_{B_i}\rangle_{Q_i}$, where $\varphi_{B_i}$ is a random angle. For the $j$-th ID-photon: Bob uses $R(-\psi_j)$ to get $|0\rangle_{P_j}$ and then measures these ID-photons. If the $m$ ID-photons are all in the state $|0\rangle$, he prepares $|\psi'_j\rangle_{P_j}$, else prepares $m$ random photons and fill them in the position of ID-photons. At last Bob sends the second message to Alice.

\item{}
After receiving the message, Alice firstly uses $R(-\psi'_j)$ to get $|0\rangle_{P_j}$ and then measures these ID-photons. If the $m$ ID-photons are all in the state $|0\rangle$, Alice can make sure that the message is coming from Bob. Then for $i$-th IF-photon: Alice uses $R(-\varphi_{A_i})$ to get state $|\varphi_{S_i}+\varphi_{C_i}+\varphi_{B_i}\rangle_{Q_i}$. For the $j$-th ID-photon: Bob uses $R(\psi''_j)$ to get $|\psi''_j\rangle_{P_j}$. At last Alice sends the third message to Bob.
If Alice find there are eavesdropping, she does not operate on IF-photon and fill random photons in the position of ID-photons. Then she sends these photons to Bob.

\item{}
After receiving Alice's message, Bob measures the ID-photons to make sure they are in the state $|\psi''_j\rangle_{P_j}$. For $i$-th IF-photons, Bob uses $R(-\varphi_{B_i}-\varphi_{C_i})$ to get $|\varphi_{S_i}\rangle_{Q_i}$. If these ID-photons are not in the state $|\psi''_j\rangle_{P_j}$, the message is not coming from Alice or has been change by attacker.
\end{enumerate}

When this protocol is used to transmit classical message, it cannot resist MIM attack without the help of preshared $\varphi_{c_i}$. Though Eve cannot know which is IF-photon and which is ID-photon, he can choose randomly from these photons, and obtain a IF-photon with non-negligible probability. Then he can carries MIM attack without being found. In detail, he interactive with Alice and Bob. When he receives the first message from Alice, he randomly select the $i$-th photon and retains it. He prepares another photon and put it in position $i$ of the message, then sends the message to Bob. When he receives the second message from Bob, he use the retained photon to replace the $i$-th photon in the message. Then he send the changed message to Alcie. In the last step, he receives the third message from Alice, he can get the $i$-th bit if the $i$-th photon is IF-photon. To resist MIM attack, the protocol can be modified as follows: the classical message to be transmitted are decomposed as a summation of $n$ bits. All the IF-photons are prepared according to these $n$ bits and are transmitted to Bob through above protocol. When Bob obtains these $n$ bits, he computes the summation of these $n$ bits and get the real classical message.


\subsection{Quantum no-key protocol for classical message transmission}\label{qnkp-cmt}
\subsubsection{A simple scheme\cite{Yangli02}}

Alice and Bob shares $\{\varphi_{C_i},i=1,\cdots,n\}$ secretly before communication. Alice wants to transmit $n$ bits classical message $x_1x_2\cdots x_n$.

At first, she prepares $n$ single-photons with the $i$-th photon in the state $|\varphi_i\rangle$, where $\varphi_i=x_i\cdot \frac{\pi}{2}$.
Then Alice and Bob communicate following the protocol in Section~\ref{sec:protocol}, and Bob obtains $|\varphi_1\rangle, \cdots, \varphi_n\rangle$, where $\varphi_i=x_i\cdot \frac{\pi}{2}$.

In the end, Bob measures the photons in bases $\{|0\rangle,|\frac{\pi}{2}\rangle\}$ one by one and gets the message $x_1x_2\cdots x_n$.

\subsubsection{Unbalance-of-Information-Source (UIS) attack\cite{Wu09}}
Wu and Yang \cite{Wu09} proposed an UIS attack to quantum no-key protocol transmitting classical messages. If $\{\varphi_{C_i}\}$ is reused for $t$ times, Eve can utilize this unbalance to attack $\{\varphi_{C_i}\}$.

Eve's strategy is: collecting all the $t$ states $|x_{i1}\cdot\frac{\pi}{2}+\varphi_{C_i}\rangle,\cdots,|x_{it}\cdot\frac{\pi}{2}+\varphi_{C_i}\rangle$ through MIM attack. Because of redundancy, the proportion of bit $0$ and bit $1$ in the information source are not equal, $p(0)=0.5+\epsilon$, $p(1)=0.5-\epsilon$, $|\epsilon|<0.5$ and $\epsilon\neq 0$. Therefore, the $t$ states can be divided into two parts whose proportion are $p(|\varphi_{C_i}\rangle)=0.5+\epsilon$ and $p(|\frac{\pi}{2}+\varphi_{C_i}\rangle)=0.5-\epsilon$, respectively. If Eve uses base $\{|0\rangle,|\frac{\pi}{2}\rangle\}$ to measure half of these states, the probability of getting $|0\rangle$ is \begin{equation}p_0=(\frac{1}{2}+\epsilon)\textrm{cos}^2\varphi_{C_i}+(\frac{1}{2}-\epsilon)\textrm{sin}^2\varphi_{C_i}=\frac{1}{2}+\epsilon\textrm{cos}2\varphi_{C_i},\end{equation} the probability of getting $|\frac{1}{2}\rangle$ is \begin{equation}p_1=(\frac{1}{2}+\epsilon)\textrm{sin}^2\varphi_{C_i}+(\frac{1}{2}-\epsilon)\textrm{cos}^2\varphi_{C_i}=\frac{1}{2}-\epsilon\textrm{cos}2\varphi_{C_i}.\end{equation}
If Eve knows the parameter $\epsilon$ of the classical message, she can obtain two angles: $\varphi_{C_{i1}}$ and $\varphi_{C_{i2}}$ ($\varphi_{C_{i1}}+\varphi_{C_{i2}}=\pi$), and one of them is $\varphi_{C_i}$. Then Eve uses the base $\{|\varphi_{C_{i1}}\rangle,|\varphi_{C_{i1}}+\frac{\pi}{2}\rangle\}$ to measure the remaining half of states,
if the proportion that they project to $|\varphi_{C_{i1}}\rangle$ is $0.5+\epsilon$,
she knows $\varphi_{C_i}=\varphi_{C_{i1}}$, otherwise, $\varphi_{C_i}=\varphi_{C_{i2}}$.

\subsubsection{A scheme using Hadamard and CNOT transformations}\label{sec:qc}
Alice prepares the base state $|x\rangle$ in a quantum register of $n$ qubits to represents a
classical message $x$ of $n$ bits, then transforms it to a superposition state via Hadamard transformations:
\begin{align}
|x_1x_2\cdots x_n\rangle\stackrel{H^{\otimes n}}{\longrightarrow}\frac{1}{\sqrt{2^{n}}} \sum_{m_{1},m_{2},...,m_{n}}(-1)^{m_1x_1+m_2x_2+...+m_nx_n}|m_1m_2\cdots m_n\rangle,
\end{align}
where $x_i(m_i)$ is the value of the $i$-th bit of message $x(m)$ . After that, Alice transmits it
with the protocol described in Section~\ref{qnkp-qc1} and \ref{qnkp-qc2}. In the end, Bob should transform the state
he has received to a base state via Hadamard transformation to get the classical message $x$.

Let us consider an example for classical message transmission: Alice needs
to transmit a $n$-bit message $x$ to Bob. Before the communication, Alice and Bob share two $n$-bit strings $s_A, s_B$. The process is as follows:
\begin{enumerate}
\item{}
Alice randomly selects $n+1$ $n$-bit numbers $k_{A_1},\cdots,k_{A_n}$ and $i$. Then Alice prepares the quantum state $|x\rangle$ to represents $x$ and transforms it to a superposition state with
Hadamard transformation
\begin{equation}|x\rangle \rightarrow \frac{1}{2^{n/2}}\sum_m(-1)^{x\cdot m}|m\rangle,\end{equation}
then performs the transformation
\begin{align}
&\frac{1}{2^{n/2}}\sum_m(-1)^{x\cdot m}|m\rangle|i\rangle \nonumber\\
\rightarrow & \frac{1}{2^{n/2}}\sum_m(-1)^{x\cdot m}|m\rangle|i\oplus m_1k_{A_1}\oplus\cdots\oplus m_nk_{A_n}\rangle,
\end{align}
where $m_1,\cdots,m_n$ are the binary string of $m$. It is an evidence that the transformation involved here can be realized by some CNOT gates (at most $n^2$ CNOT gates). After the computation, Alice sends the $2n$-qubit state to Bob.

\item{}
Bob randomly selects $n+1$ $n$-bit numbers $k_{B_1},\cdots,k_{B_n}$ and $j$, then computes the transformation:
\begin{align}
&\frac{1}{2^{n/2}}\sum_m(-1)^{x\cdot m}|m\rangle|i\oplus m_1k_{A_1}\oplus\cdots\oplus m_nk_{A_n}\rangle|j\rangle \nonumber\\
\rightarrow &\frac{1}{2^{n/2}}\sum_m(-1)^{x\cdot m}|m\rangle|i\oplus m_1k_{A_1}\oplus\cdots\oplus m_nk_{A_n}\oplus s_B\rangle\otimes\nonumber\\
&\otimes|j\oplus m_1k_{B_1}\oplus\cdots\oplus m_nk_{B_n}\rangle,
\end{align}
and then Bob sends this $3n$-qubit state back to Alice.

\item{}
Alice computes
\begin{align}
&\frac{1}{2^{n/2}}\sum_m(-1)^{x\cdot m}|m\rangle|i\oplus m_1k_{A_1}\oplus\cdots\oplus m_nk_{A_n}\oplus s_B\rangle\otimes\nonumber\\
&\otimes|j\oplus m_1k_{B_1}\oplus\cdots\oplus m_nk_{B_n}\rangle \nonumber\\
\rightarrow & \frac{1}{2^{n/2}}\sum_m(-1)^{x\cdot m}|m\rangle|0\rangle|j\oplus m_1k_{B_1}\oplus\cdots\oplus m_nk_{B_n}\oplus s_A\rangle,
\end{align}
and then measures the $i+1\sim2i$-th qubit to check whether they are in state $|0\rangle$. Then Alice sends the $2n$-qubit state $\frac{1}{2^{n/2}}\sum_m(-1)^{x\cdot m}|m\rangle|j\oplus m_1k_{B_1}\oplus\cdots\oplus m_nk_{B_n}\oplus s_A\rangle$ to Bob.

\item{}
Bob computes
\begin{align}
&\frac{1}{2^{n/2}}\sum_m(-1)^{x\cdot m}|m\rangle|j\oplus m_1k_{B_1}\oplus\cdots\oplus m_nk_{B_n}\oplus s_A\rangle\nonumber\\
\rightarrow &\frac{1}{2^{n/2}}\sum_m(-1)^{x\cdot m}|m\rangle.
\end{align}
Then he does an Hadamard transformation to the $n$-qubit state and get $|x\rangle$.
\end{enumerate}

This protocol is also a practical one, since all computation involved can be implemented with Hadamard and single-level CNOT gates. It is worth to be investigated that whether the local random numbers $k_{A_1},\cdots,k_{A_n},i,k_{B_1},\cdots$,\\$k_{B_n},j$ can protect $s_A$ and $s_B$.

\section{QNK protocols based on quantum perfect encryption}\label{sec:qnkp-qpe}
\subsection{Quantum perfect encryption}
Suppose a set of operations $U_k,k=1,2,\cdots,N$ is open, each element $U_k$ is $2^n\times2^n$ unitary matrix. Let the cipher state of a n-qubit quantum message $\rho$ is $\rho_c$. In the encryption stage, $U_k$ is applied to the quantum state, where $k$ is a secret key, each $k$ is chosen with probability $p_k$ for Alice.
\begin{equation}
\rho_c=U_k\rho U_k^{\dagger}.
\end{equation}
And in the decryption stage, $U_k^{\dagger}$ is applied to the cipher state $\rho_c$,
\begin{equation}
\rho=U_k^{\dagger}\rho_c U_k.
\end{equation}
Quantum perfect encryption is defined as \cite{Boykin00}: for every input state $\rho$, the output state is a totally mixed state, that is
\begin{equation}
\sum_k p_k U_k\rho U_k^{\dagger}=\frac{I}{2^n}.
\end{equation}
\cite{Boykin00} constructs one perfect encryption by choosing $p_k=\frac{1}{2^{2n}}$, $U_k=X^{\alpha}Z^{\beta} (\alpha,\beta\in\{0,1\}^n)$. Via defining the inner product of two matrices $M_1$ and $M_2$ as $Tr(M_1M_2^{\dagger})$, the set of all $2^n\times2^n$ matrices can be  regarded as an inner product space. Then it can be proven that the set of $2^{2n}$ unitary matrices $\{X^{\alpha}Z^{\beta}\}$ forms an complete orthonormal basis. Any message state $\rho$ can be expanded as
\begin{equation}
\rho=\sum_{\alpha,\beta}a_{\alpha,\beta}X^{\alpha}Z^{\beta},
\end{equation}
where $a_{\alpha,\beta}=tr(\rho Z^{\beta}X^{\alpha})/{2^n}$. Boykin and Roychowdhury prove that their construction is perfect.

\subsection{Quantum perfect encryption based on generalized quantum commutative transformations}\label{sec:qpe-qct}
We propose a quantum perfect encryption scheme based on a set of generalized quantum commutative transformation.
Given two $2\times 2$ unitary transformations $U_1$ and $U_2$, which satisfy the following relation $$U_1U_2=-U_2U_1.$$

We choose $p_k=\frac{1}{2^{2n}}$, $U_k=U_1^{\alpha}U_2^{\beta}, k=(\alpha,\beta)$, where $\alpha,\beta\in\{0,1\}^n$. In order to satisfy the requirement of quantum perfect encryption, the unitary transformations $U_1$ and $U_2$ should satisfy: $\{U_1,U_2,U_1U_2,I\}$ is an complete orthonormal basis. That is, the four unitary matrixes are mutually orthonormal. Thus, we can conclude the following formulas:
\begin{enumerate}
\item{}
$0=(U_1U_2,I)=tr(U_2^\dagger U_1^\dagger I)=tr(U_2^\dagger U_1^\dagger)=(tr(U_1U_2))^*$, that is $tr(U_1U_2)=0$.
\item{}
$0=(U_1,U_2)=tr(U_1^\dagger U_2)$.
\item{}
$0=(U_1,U_1U_2)=tr(U_1^\dagger U_1U_2)=tr(U_2)$.
\item{}
$0=(U_2,U_1U_2)=tr(U_2^\dagger U_1U_2)=tr(U_1U_2U_2^\dagger)=tr(U_1)$.
\item{}
$0=(U_1,I)=tr(U_1^\dagger)$, that is $tr(U_1)=0$.
\item{}
$0=(U_2,I)=tr(U_2^\dagger)$, that is $tr(U_2)=0$.
\end{enumerate}
Therefore, the unitary matrixes $U_1$ and $U_2$ should satisfy the conditions $tr(U_1)=tr(U_2)=tr(U_1U_2)=tr(U_1^\dagger U_2)=0$ and $U_1U_2=-U_2U_1$.

Similar to the security proof of $\{p_k=\frac{1}{2^{2n}},U_k=X^{\alpha}Z^{\beta}, k=(\alpha,\beta),\alpha,\beta\in\{0,1\}^n\}$ in \cite{Boykin00}, we get the following results.

{\bf Proposition 6:} $\{p_k=\frac{1}{2^{2n}},U_k=U_1^{\alpha}U_2^{\beta}, k=(\alpha,\beta),\alpha,\beta\in\{0,1\}^n\}$ is a quantum perfect encryption.

{\bf Proof:}
Because $\{U_1^\alpha U_2^\beta,\alpha,\beta\in\{0,1\}^n\}$ is a complete orthonormal basis, any $n$-qubit state $\rho$ can be represented as a linear combination of these $2^{2n}$ unitary matrixes: $$\rho=\sum_{\alpha,\beta} a_{\alpha,\beta} U_1^\alpha U_2^\beta,$$
where $a_{\alpha,\beta}=tr(\rho U_2^\beta U_1^\alpha)/{2^n}$. Then,

\begin{eqnarray*}
\sum_k p_k U_k\rho U_k^\dagger &=& \frac{1}{2^{2n}}\sum_{\gamma,\delta}U_1^\gamma U_2^\delta \rho U_2^\delta U_1^\gamma \\
&=& \frac{1}{2^{2n}}\sum_{\alpha,\beta}a_{\alpha,\beta}\sum_{\gamma,\delta}U_1^\gamma U_2^\delta U_1^\alpha U_2^\beta U_2^\delta U_1^\gamma.
\end{eqnarray*}
From $U_1U_2=-U_2U_1$,we have $U_2^\delta U_1^\alpha=(-1)^{\alpha\cdot\delta}U_1^\alpha U_2^\delta$. Thus, the above expression is equal to
\begin{eqnarray*}
&& \frac{1}{2^{2n}}\sum_{\alpha,\beta}a_{\alpha,\beta}\sum_{\gamma,\delta}(-1)^{\alpha\cdot\delta}U_1^\alpha U_1^\gamma U_2^\delta (-1)^{\beta\cdot\gamma} U_2^\delta U_1^\gamma U_2^\beta\\
&=& \frac{1}{2^{2n}}\sum_{\alpha,\beta}a_{\alpha,\beta}\sum_{\gamma,\delta}(-1)^{\alpha\cdot\delta}(-1)^{\beta\cdot\gamma}U_1^\alpha U_2^\beta.
\end{eqnarray*}
Because $\frac{1}{2^n}\sum_{\gamma\in\{0,1\}^n}(-1)^{\beta\cdot\gamma}=\delta_{\beta,0}$, the above formula is equal to
$$
\sum_{\alpha,\beta}a_{\alpha,\beta}\delta_{\alpha,0}\delta_{\beta,0}U_1^\alpha U_2^\beta = a_{00}I=\frac{tr(\rho)}{2^n}I=\frac{I}{2^n}.
$$
So, the scheme is a quantum perfect encryption.$\Box$

There are many special cases satisfying the conditions of $U_1$ and $U_2$, such as $X$ and $Y$, $Y$ and $H$, $X$ and $Z$. Thus, the following examples are all quantum perfect encryptions.
\begin{enumerate}
\item{}
PQC1:$\{p_k=\frac{1}{2^{2n}},U_k=X^{\alpha}Y^{\beta}, k=(\alpha,\beta),\alpha,\beta\in\{0,1\}^n\}$.
\item{}
PQC2:$\{p_k=\frac{1}{2^{2n}},U_k=Y^{\alpha}H^{\beta}, k=(\alpha,\beta),\alpha,\beta\in\{0,1\}^n\}$.
\item{}
PQC3:$\{p_k=\frac{1}{2^{2n}},U_k=X^{\alpha}Z^{\beta}, k=(\alpha,\beta),\alpha,\beta\in\{0,1\}^n\}$. This is just the case introduced in \cite{Boykin00}.
\end{enumerate}

\subsection{Quantum no-key protocol based on quantum perfect encryption}\label{qnkp-qpe}
For any two unitary transformation $U_k=U_1^\alpha U_2^\beta$ and $U_l=U_1^\gamma U_2^\delta$, we have
\begin{eqnarray*}
U_kU_l &=& (U_1^\alpha U_2^\beta)(U_1^\gamma U_2^\delta) \\
&=& U_1^\alpha (U_2^\beta U_1^\gamma) U_2^\delta \\
&=& U_1^\alpha (-1)^{\beta\cdot\gamma}U_1^\gamma U_2^\beta U_2^\delta \\
&=& (-1)^{\beta\cdot\gamma}U_1^\gamma (U_1^\alpha U_2^\delta) U_2^\beta \\
&=& (-1)^{\beta\cdot\gamma+\alpha\cdot\delta}(U_1^\gamma U_2^\delta)(U_1^\alpha U_2^\beta)=(-1)^{\beta\cdot\gamma+\alpha\cdot\delta}U_l U_k,
\end{eqnarray*}
where $k=(\alpha,\beta)$, $l=(\gamma,\delta)$.

Thus, according to Proposition 4, the following protocol constructed from the PQC:$\{p_k=\frac{1}{2^{2n}},U_k=U_1^{\alpha}U_2^{\beta}, k=(\alpha,\beta),\alpha,\beta\in\{0,1\}^n\}$ holds.
\begin{enumerate}
\item{}
Alice randomly selects $\alpha_A,\beta_A \in \{0,1\}^n$, and encrypts $\rho$ with $U_1^{\alpha_A}U_2^{\beta_A}$, and sends Bob $\rho_1=U_1^{\alpha_A}U_2^{\beta_A}\rho (U_1^{\alpha_A}U_2^{\beta_A})^{\dagger}$.
\item{}
Bob randomly selects $\alpha_B,\beta_B \in \{0,1\}^n$, and encrypts $\rho_1$ with $U_1^{\alpha_B}U_2^{\beta_B}$, and sends Alice $\rho_2=U_1^{\alpha_B}U_2^{\beta_B}\rho_1 (U_1^{\alpha_B}U_2^{\beta_B})^{\dagger}$.
\item{}
Alice decrypts $\rho_2$ with $(U_1^{\alpha_A}U_2^{\beta_A})^{\dagger}$ and sends Bob $\rho_3=(U_1^{\alpha_A}U_2^{\beta_A})^{\dagger}\rho_2 U_1^{\alpha_A}U_2^{\beta_A}$.
\item{}
Bob decrypts $\rho_3$ with $(U_1^{\alpha_B}U_2^{\beta_B})^\dagger$ to recover $\rho$.
\end{enumerate}

Each of the three PQCs listed in Section~\ref{sec:qpe-qct} can be used in the above protocol. If we choose $\{p_k=\frac{1}{2^{2n}},U_k=Y^{\alpha}H^{\beta},\alpha,\beta\in\{0,1\}^n\}$ for the protocol. Then the protocol is as follows:

\begin{enumerate}
\item{}
Alice encrypts $\rho$ with $Y^{\alpha_A}H^{\beta_A}$, and sends Bob $\rho_1=Y^{\alpha_A}H^{\beta_A}\rho H^{\beta_A}Y^{\alpha_A}$.
\item{}
Bob encrypts $\rho_1$ with $Y^{\alpha_B}H^{\beta_B}$ and sends Alice $\rho_2=Y^{\alpha_B}H^{\beta_B}\rho_1 H^{\beta_B}Y^{\alpha_B}$.
\item{}
Alice decrypts $\rho_2$ with $H^{\beta_A}Y^{\alpha_A}$ and sends Bob $\rho_3=H^{\beta_A}Y^{\alpha_A}\rho_2 Y^{\alpha_A}H^{\beta_A}$.
\item{}
Bob decrypts $\rho_3$ with $H^{\beta_B}Y^{\alpha_B}$ to recover $\rho$.
\end{enumerate}

It can be seen that these protocols can also transmit classical information after the classical information being encoded into computational basis state.

{\bf Remark 4:} (a) When we choose the PQC $\{p_k=\frac{1}{2^{2n}},U_k=X^{\alpha}Z^{\beta},\alpha,\beta\in\{0,1\}^n\}$ for the quantum no-key protocol, it is unsafe to transmit classical information. Because after the classical bits being encoded into computational basis state, it will stay in computational basis state during the exchange in the protocol. Thus the attacker can measure the cipher in the basis $\{|0\rangle,|1\rangle\}$ without breaking it. And because the three ciphers transmitted between Alice and Bob is $X^{\alpha_A}Z^{\beta_A}|m\rangle$,$X^{\alpha_B}Z^{\beta_B} X^{\alpha_A}Z^{\beta_A}|m\rangle$,$X^{\alpha_B}Z^{\beta_B}|m\rangle$ ($m$ is the classical message), measuring the three ciphers can achieve the three strings $\alpha_A\oplus m, \alpha_B\oplus\alpha_A\oplus m, \alpha_B\oplus m$. The attacker can computes $\alpha_B$ from the first string and the second string. Then he can computes the message $m$ from the value of $\alpha_B$ and the third string.

(b) When we choose the PQC $\{p_k=\frac{1}{2^{2n}},U_k=X^{\alpha}Y^{\beta},\alpha,\beta\in\{0,1\}^n\}$ for the quantum no-key protocol, it is also unsafe to transmit classical information for the same reason. In this case, the three ciphers transmitted between Alice and Bob is $X^{\alpha_A}Y^{\beta_A}|m\rangle$,$X^{\alpha_B}Y^{\beta_B} X^{\alpha_A}Y^{\beta_A}|m\rangle$,$X^{\alpha_B}Y^{\beta_B}|m\rangle$ ($m$ is the classical message), measuring the three ciphers can achieve the three strings $\alpha_A\oplus\beta_A\oplus m, \alpha_B\oplus\beta_B\oplus\alpha_A\oplus\beta_A\oplus m, \alpha_B\oplus\beta_B\oplus m$. The attacker can computes $\alpha_B\oplus\beta_B$ from the first string and the second string. Then he can computes the message $m$ from the value of $\alpha_B\oplus\beta_B$ and the third string.

Through the above remark, we know that it is better to choose the PQC $\{p_k=\frac{1}{2^{2n}},U_k=Y^{\alpha}H^{\beta}, k=(\alpha,\beta),\alpha,\beta\in\{0,1\}^n\}$ for the quantum no-key protocol. By using $Y^{\alpha}H^{\beta}$ in the protocol, the message is just being encoded into the conjugate coding, and the flaw stated in the above remark disappears.

\section{Quantum no-key protocol based on Boolean function computing}\label{sec:qnkp-qc}
\subsection{Protocol for quantum message transmission\cite{Yang03}}\label{qnkp-qc1}
A quantum message is a sequence of pure states:
\begin{eqnarray}
M_{k}^{(n)}=\{\sum_m\alpha_{m}^{(i)}|m\rangle | i=1,2,...n\},
\end{eqnarray}
where $ m=(m_{1},m_{2},...,m_{k})\in \{0,1\}^k$.
Let us consider the secure transmission of a pure state $ \sum_m\alpha_m|m\rangle $.
Here "secure" means 1) Eve cannot get the state even when she has controlled the channel;
2) Bob can verify that the state really comes from Alice;
3) Alice can verify that the receiver is Bob; 4) Bob know whether the state has been changed in the channel.
 These are so called encryption, identification and authentication of message.

Because the two unitary transformations
$$U_A:\sum_m \alpha_{m}|m\rangle|0\rangle|0\rangle\rightarrow\sum_m \alpha_{m}|m\rangle|F_{A}(m)\rangle|0\rangle $$ and
$$U_B:\sum_m \alpha_{m}|m\rangle|0\rangle|0\rangle\rightarrow\sum_m \alpha_{m}|m\rangle|0\rangle|F_{B}(m)\rangle$$
are commutative, according to Proposition 4, we can construct a quantum no-key protocol using this kind of unitary transformations.
Here is the basic encryption protocol
for quantum message without authentication:
\begin{enumerate}
\item{}
Alice randomly chooses a $n$-dimensional Boolean function
 \begin{eqnarray}
 F_{A}(x)=(f_{A}^{1}(x), f_{A}^{2}(x)..., f_{A}^{n}(x))
\end{eqnarray}
where $f_A^i(x):\{0,1\}^k\rightarrow\{0,1\}$, and performs an unitary transformation as
below:
\begin{eqnarray}
\sum_m \alpha_{m}|m\rangle|0\rangle\rightarrow\sum_m \alpha_{m}|m\rangle|F_{A}(m)\rangle
\end{eqnarray}
then sends the state to Bob.
\item{}
Bob chooses his Boolean function $F_{B}(x)$ independently and randomly, and performs an unitary transformation as
below:
\begin{eqnarray}
\sum_m \alpha_{m}|m\rangle|F_{A}(m)\rangle|0\rangle\rightarrow\sum_m \alpha_{m}|m\rangle|F_{A}(m)\rangle|F_{B}(m)\rangle
\end{eqnarray}
then sends it back to Alice.

\item{}
Alice performs the following transformation:

\begin{equation}
\begin{split}
&\sum_m \alpha_{m}|m\rangle|F_{A}(m)\rangle|F_{B}(m)\rangle \\
\rightarrow &\sum_m \alpha_{m}|m\rangle|F_{A}(m)\oplus F_{A}(m)\rangle |F_{B}(m)\rangle \\
= &\sum_m \alpha_{m}|m\rangle|0\rangle|F_{B}(m)\rangle,
\end{split}
\end{equation}
and sends $\sum_m \alpha_{m}|m\rangle|F_{B}(m)\rangle$ to Bob.

\item{}
Bob does the same computation with his function $ F_{B}(x) $
\begin{equation}
\begin{split}
&\sum_m \alpha_{m}|m\rangle|F_{B}(m)\rangle \\
\rightarrow &\sum_m \alpha_{m}|m\rangle|F_{B}(m)\oplus F_{B}(m)\rangle
= \sum_m \alpha_{m}|m\rangle|0\rangle,
\end{split}
\end{equation}
then gets the quantum message $\sum_m\alpha_m|m\rangle$.
\end{enumerate}

\subsection{Improved protocol with personal identification\cite{Yang03}}\label{qnkp-qc2}
The protocol in Section \ref{qnkp-qc1} cannot defend MIM attack, and we can modify it by adding personal identification. Suppose Alice and Bob preshare identification keys $s_{A}$ and $s_{B}$ , where $s_{A}$ and $s_{B}$
are Boolean functions. The modified protocol is as follows:

\begin{enumerate}
\item{}
Alice prepares the state as below:
\begin{eqnarray}
\sum_{m}\alpha_{m}|m\rangle|0\rangle|0\rangle & \rightarrow & \sum_{m}\alpha_{m}|m\rangle|F_A(m)\rangle|0\rangle \nonumber\\
& \rightarrow & \sum_{m}\alpha_{m}|m\rangle|F_A(m)\rangle|s_{A}(m)\rangle,
\end{eqnarray}
and sends it to Bob.

\item{}
Bob performs the following transformation
\begin{equation}
\rightarrow\sum_{m}\alpha_{m}|m\rangle|F_{A}(m)\rangle|0\rangle
\end{equation}
and verifies that the message is really coming from Alice via measuring the third register, and then performs the following transformation:
\begin{equation}
\begin{split}
\rightarrow &\sum_{m}\alpha_{m}|m\rangle|F_{A}(m)\rangle|F_{B}(m)\rangle \\
\rightarrow &\sum_{m}\alpha_{m}|m\rangle|F_{A}(m)\oplus s_{B}(m)\rangle|F_{B}(m)\rangle,
\end{split}
\end{equation}
and sends it back to Alice.

\item{}
Alice transforms the state and verifies that the quantum message is really coming back
from Bob:
\begin{equation}
\begin{split}
\sum_{m}\alpha_{m}|m\rangle|F_{A}(m)\oplus s_{B}(m)\rangle|F_{B}(m)\rangle\rightarrow &\sum_{m}\alpha_{m}|m\rangle|F_{A}(m)\rangle|F_{B}(m)\rangle\\
\rightarrow &\sum_{m}\alpha_{m}|m\rangle|0\rangle|F_{B}(m)\rangle.
\end{split}
\end{equation}
If the second quantum register is in state $|0\rangle$ , Alice believes that it really
comes from Bob, otherwise she stops the protocol. When Eve pretend to be Alice to communicate with Bob, she can substitute the
second register with one in the state $F_{E}(m)\rangle$ , but she cannot transform the third
register into $|F_{B}(m)\oplus s_{A}(m)\rangle$ if we choose $s_{A}\neq s_{B}$. Finally Alice transforms the
state to $\sum_{m}\alpha_{m}|m\rangle|F_{B}(m)\oplus s_{A}(m)\rangle$, and sends it to Bob again.

\item{}
Bob transforms the state as below to get the message,
\begin{equation}
\begin{split}
\sum_{m}\alpha_{m}|m\rangle|F_{B}(m)\oplus s_{A}(m)\rangle\rightarrow &\sum_{m}\alpha_{m}|m\rangle|F_{B}(m)\rangle \\
\rightarrow &\sum_{m}\alpha_{m}|m\rangle|0\rangle,
\end{split}
\end{equation}
and verifies Alice's legitimacy via measuring the
second register.
\end{enumerate}

In this protocol, $F_{A}$ and $F_{B}$ are used to protect $s_{A}$ and $s_{B}$.

\subsection{Protocol with ancillary quantum state}\label{p-aqs}
We define an unitary transformation $U_f$ as follows:
\begin{equation}\label{eq:U_f}
U_f:\sum_m\alpha_m|m\rangle \rightarrow \sum_m\alpha_m(-1)^{f(m)}|m\rangle,
\end{equation}
where $f$ is a boolean function $f:\{0,1\}^n\rightarrow\{0,1\}$. The unitary transformation $U_f$ can be implemented with the help of an ancillary qubit as
\begin{eqnarray*}
&&|m\rangle\frac{|0\rangle-|1\rangle}{\sqrt{2}} \\
&\stackrel{U_f}{\longrightarrow}& |m\rangle\frac{|f(m)\rangle-|f(m)\oplus1\rangle}{\sqrt{2}}=(-1)^{f(m)}|m\rangle\frac{|0\rangle-|1\rangle}{\sqrt{2}}
\end{eqnarray*}

It can be seen that $U_{f_1}U_{f_2}=U_{f_2}U_{f_1}$, where $U_{f_1}$ and $U_{f_2}$ are defined as Eq.~(\ref{eq:U_f}).
A protocol for transmitting $n$-qubit state $\sum_m\alpha_m|m\rangle$ based on them is as follows. 

Suppose a set of Boolean functions $\{f_i\}$ is shared by Alice and Bob.

\begin{enumerate}
\item{}
Alice randomly selects a function $f_A$, and performs $U_{f_A}$ on $\sum_m\alpha_m|m\rangle$,
\begin{equation}
\sum_m\alpha_m|m\rangle \rightarrow \sum_m\alpha_m(-1)^{f_A(m)}|m\rangle,
\end{equation}
and then sends it to Bob.

\item{}
Bob randomly selects a function $f_B$, and performs $U_{f_B}$ as follows,
\begin{equation}
\sum_m\alpha_m(-1)^{f_A(m)}|m\rangle \rightarrow \sum_m\alpha_m(-1)^{f_A(m)+f_B(m)}|m\rangle,
\end{equation}
and then sends the state to Alice.

\item{}
Alice performs $U_{f_A}$ again, then
\begin{equation}
\sum_m\alpha_m(-1)^{f_A(m)+f_B(m)}|m\rangle \rightarrow \sum_m\alpha_m(-1)^{f_B(m)}|m\rangle,
\end{equation}
and then sends it to Bob.

\item{}
Bob performs $U_{f_B}$ again,
\begin{equation}
\sum_m\alpha_m(-1)^{f_B(m)}|m\rangle \rightarrow \sum_m\alpha_m|m\rangle,
\end{equation}
then he gets the quantum message $\sum_m\alpha_m|m\rangle$.
\end{enumerate}


\section{Discussions}
Quantum no-key protocols without personal identification cannot resist MIM attack. In order to resist the MIM attack, personal identification must be added into protocols. In Section~\ref{sec:qnkp-pi}, we describe a general way to add personal identification into a quantum no-key protocol.

The protocol in Section~\ref{sec:protocol-qmt} have no identification function. A set of preshared $\varphi_{C_i}$ is used for personal identification in Section~\ref{sec:protocol}, but Alice and Bob cannot identify each other in every pass during the three times of interactive. In the protocol described in Section~\ref{ps-mi}, some qubits are used only for identification. In this protocol, Alice and Bob use the preshared personal information to identify each other in each pass of interactive, then it satisfies the way of identification introduced in Section~\ref{sec:qnkp-pi}. A protocol in Section~\ref{qnkp-qc2} also adopts this kind of identification. It can be seen that the identification can be added into the protocol in Section~\ref{p-aqs} in the same way.

If Alice and Bob identify each other in each time of interactive, the four operations performed successively by Alice and Bob must satisfy some relations. For instance, in the framework presented in Section~\ref{sec:qnkp-pi}, the two operators $U_k(s_A)$ and $V_l'(s_A)$ must satisfy the following relation:
$$V_l'(s_A)U_k(s_A)=U_M(k,l)\otimes I_A,\forall k,l,s_A,$$
where operators $U_k(s_A)$ and $V_l'(s_A)$ are both relative to the preshared personal information $s_A$. This formula means that the operator $V_l'(s_A)$ can remove the change of identification qubits caused by operator $U_k(s_A)$.

Preshare personal identities $s_A,s_B$ is necessary for identifying each other, so the privacy of $s_A,s_B$ is important to the security of the protocol. An essential problem of QNK protocol is whether $s_A,s_B$ can be reused under the protection of those local random numbers of Alice and Bob.

There are three times of transmission of quantum ciphers in a QNK protocol. Consider of the relations among these three ciphers, it is necessary to investigate whether there exists a kind of attack making use of these relations. For this kind of interactive protocol, how to define its security is still an open problem.

Generally, the protocols in this paper can be used to transmit both classical and quantum messages. While Alice transmits a classical $n$-bit message $x$ to Bob, she can encode the classical message into a quantum state ( one of computational basis states), and perform Hadamard transformations $H^{\otimes n}$ on this quantum state, and then transmit the quantum state to Bob. However, some quantum message oriented protocols are not secure when transmitting classical message (see the discussions in Section~\ref{ps-mi},~\ref{qnkp-cmt} and ~\ref{qnkp-qpe}). Furthermore, when the protocols in this paper are used to transmit classical message, whether it can resist the UIS attack described in Section~\ref{qnkp-cmt} needs further investigation.

Some practical quantum no-key protocols are described in Section~\ref{sec:practical}. One of these protocols involving only rotations of single-photons can be implemented with current techniques. It is believed that protocols based on single-qubit rotation and single-level CNOT gates may also be implemented in the near future.

\section{Conclusions}
A theoretical framework of QNK protocol is proposed. Some practical QNK protocols are reviewed and a new protocol is presented. QNK protocols based on the scheme of quantum perfect encryption are proposed. Protocols based on Boolean function computing are also discussed. Some of the protocols in this paper are secure against man-in-the-middle attack, beyond computational hypothesis.

\section*{Acknowledgements}
This work was supported by the National Natural Science Foundation of
China under Grant No. 61173157.


\begin{thebibliography}{99}
\bibitem{Boykin00}
P. Boykin and V. Roychowdhury, Optimal Encryption of Quantum Bits, Arxiv preprint quant-ph/0003059.

\bibitem{Ambainis00}
A. Ambainis, et al, Private quantum channels, 41st Annual Symposium on Foundations of Computer Science, Proceedings: 547-553, 2000.

\bibitem{Nayak07}
A. Nayak and P. Sen, Invertible quantum operations and perfect encryption of quantum states, Quantum Information \& Computation 7(1-2): 103-110, 2007.

\bibitem{Yang03}
L. Yang, Quantum public-key cryptosystem based on classical NP-complete problem, Arxiv preprint quant-ph/0310076.

\bibitem{Yang10}
L. Yang, et al, Quantum public-key cryptosystems based on induced trapdoor one-way transformations, Arxiv preprint arXiv:1012.5249.

\bibitem{Menezes97}
A. J. Menezes, P. C. van Oorschot, and S. A. Vanstone, Handbook of Applied Cryptography, CRC Press, Boca Raton,1997

\bibitem{Yangli02a}
L. Yang, L. A. Wu, Transmit Classical and Quantum Information Secretly. arXiv:quant-ph/0203089.

\bibitem{Yangli02}
L. Yang, L. A. Wu, and S. H. Liu, Proc. SPIE, 4917(2002), 106-111.

\bibitem{Yangli03}
L. Yang, Quantum no-key protocol for direct and secure transmission of quantum and classical messages. Quant-ph/ 0309200, 28 Sep 2003.

\bibitem{Yangli06}
L. Yang and L. Hu, Quantum no-key protocol with inherent identification, Proc. SPIE Vol. 6305, pp. 63050J (2006).

\bibitem{Kanamori05}
Y. Kanamori, S. M. Yoo and Mohammad, A Quantum No-Key Protocol for Secure Data Communication, 43rd ACM SE Conference, March 18-20, 2005

\bibitem{Kye05}
W. H. Kye, C. M. Kim, M. S. Kim and Y. J. Park, Quantum Key Distribution with Blind Polarization Bases, Phys.Rev.Lett. 95 (4), 2005, 040501.

\bibitem{Subhash07}
S. Kak, A Three-Stage Quantum Cryptography Protocol, Foundations of Physics Letters, Vol. 19, No. 3, June 2006.

\bibitem{Wu09}
Y. Wu and L. Yang, Practical quantum no-key protocol with identification. IAS 2009: 540-543, IEEE Computer Society.

\bibitem{Beige01}
A. Beige, et al. Secure communication with a publicly known key. Acta physica Polonica. A 101(3): 357-368. (see also arXiv:quant-ph/0101066).

\bibitem{Bostrom02}
K. Bostr\"om and T. Felbinger. Deterministic secure direct communication using entanglement. Physical Review Letters 89(18): 187902. 2002.

\bibitem{Deng2003}
F. G. Deng, G. L. Long and X. S. Liu. Two-step quantum direct communication protocol using the Einstein-Podolsky-Rosen pair block. Physical Review A 68(4): 042317. 2003.

\bibitem{Deng2004}
F. G. Deng and G. L. Long. Secure direct communication with a quantum one-time pad. Physical Review A 69(5): 52319. 2004.

\end{thebibliography}
\end{document}